# A simple method to extend COBRApy dynamic FBA protocol to the metabolic simulation of communities.

Domenico L. Gatti

*Dept. of Biochemistry, Microbiology and Immunology, Wayne State University School of Medicine, 540 E. Canfield Av, Detroit, 48201, USA*

*Introduction.*

Over the past decade, constraint-based modelling (CBM) has emerged as an important approach to obtain information about physiological and perturbed metabolic states at genome-scale. *Flux balance analysis* (FBA) [1, 2] is based on stoichiometric cell models that mathematically represent the biochemical reactions in a metabolic network. The essential information required to construct a stoichiometric model is a list of participating metabolites, a list of intracellular reactions involving these metabolites, and the stoichiometric coefficients for every reaction. Each intracellular metabolite is assumed to maintain a stable concentration over time as the fluxes producing the metabolite are balanced by the fluxes consuming it. *Steady-state* mass balances can thus be represented as a matrix equation of the form:

$$\boldsymbol{A}\boldsymbol{v} = \frac{d\boldsymbol{z}}{dt} = \boldsymbol{0}$$

where $\boldsymbol{A}$ is the stoichiometric matrix with *m* rows corresponding to the metabolites and *n* columns corresponding to the reactions, $\boldsymbol{v}$ is a vector of fluxes (with units of mmol/gDW/h), and $\boldsymbol{z}$ is a vector of metabolites concentrations (with units of mmol/L). The matrix entry $a_{ij}$ is the stoichiometry of the $i^{th}$ species participating in the $j^{th}$ reaction.

The objective of FBA is to solve for the fluxes $\boldsymbol{v}$. For most stoichiometric models of practical interest, the metabolic network contains more unknown fluxes than intracellular species, and the linear system is underdetermined. Typically, fluxes are computed by solving a *linear programming* (LP) problem formulated under the assumption that the cell utilizes available resources to maximize an *objective*, most often, growth. The growth rate $\mu$ (with units of inverse time $h^{-1}$) is calculated as the weighted sum of the fluxes contributing to the biomass formation. The LP problem to be solved has the form:

$$\max_{\boldsymbol{v}(t)} \mu = \boldsymbol{w}^{\boldsymbol{T}}\boldsymbol{v}$$
$$s.t.$$

$$\boldsymbol{v}_{\boldsymbol{min}} \leq \boldsymbol{v}(t) \leq \boldsymbol{v}_{\boldsymbol{max}}$$

On the other hand, the steady-state assumption in FBA is not optimal for biological systems where the environment changes rapidly over time. Dynamic FBA (dFBA) [3] extends the capabilities of FBA, which analyzes metabolic networks at a single time point, to track changes in metabolite concentrations and fluxes over time in response to environmental conditions and cellular processes.

dFBA models assume intracellular reactions are much faster than extracellular ones, so that intracellular metabolites can be considered at *quasi*-steady state. Therefore, the growth rate $\mu$, and the intracellular fluxes $\boldsymbol{v}$ are computed through solution of the classical FBA problem via LP. FBA is then expanded to dFBA by adding mass balances for the extracellular metabolites and kinetics for their uptake/release. Rather than specifying constant transport rates, extracellular substrate concentrations $\boldsymbol{S}$ and product concentrations $\boldsymbol{P}$ are used to calculate time-varying substrate uptake rates $\boldsymbol{v_s}$ and product secretion rates $\boldsymbol{v_p}$. Because of transport limitations, these represent maximum possible rates and are incorporated as *upper bounds* on the calculated rates. The extracellular concentrations are computed by solving extracellular balance equations for the biomass concentration $X$ (in units of gDW/L) and the substrate and product concentrations (in units of mmol/L) given the growth and secretion rates obtained by LP.

$$\frac{d\boldsymbol{S}}{dt} = -\boldsymbol{v_s}X$$

$$\frac{d\boldsymbol{P}}{dt} = \boldsymbol{v_p}X$$

$$\frac{dX}{dt} = \mu X$$

where $\boldsymbol{S}$, $\boldsymbol{P}$, and X are the substrate, product, and biomass concentrations at time $t$. Although the cell is assumed to maintain an intracellular steady state, all the intracellular and extracellular variables are time varying.

Converting the *differential equations* for the time evolution of extracellular metabolites and biomass into *difference equations* over small time intervals results in a set of algebraic equations that can be solved using standard ordinary differential equation (ODE) solvers. This approach is referred to as the *static optimization approach* (SOA) for dFBA. In this method, the time period is divided into N intervals of length $\Delta T$.

$$\boldsymbol{S}(t + \Delta T) = \boldsymbol{S}(t) + \frac{d\boldsymbol{S}}{dt}\Delta T$$

$$\boldsymbol{P}(t + \Delta T) = \boldsymbol{P}(t) + \frac{d\boldsymbol{P}}{dt}\Delta T$$

$$X(t + \Delta T) = X(t) + \frac{dX}{dt}\Delta T$$

dFBA starts with an initial set of metabolite concentrations and a flux distribution obtained from FBA. At each time step, it updates the extracellular metabolite concentrations based on uptake and secretion. This process is repeated iteratively from $t_0$ to $t_f$ simulating the dynamic behavior of the system over time, until the objective function becomes infeasible (e.g. by a lack of nutrients), or the final time point is reached.

In its basic implementation, this procedure is a variation of the *Forward Euler* method, and therefore it is an *explicit method* that provides 1st order accuracy $O(\Delta T)$. Since the anticipated error is the same order of magnitude as the chosen stepsize $\Delta T$, a very small $\Delta T$ must be taken to ensure the result is as correct as possible. Higher-order accuracy than the basic SOA described above, can be obtained using more advanced *explicit* or *implicit Runge-Kutta* methods [4] to solve the ODEs that arise in the dFBA framework.

The CBM package COBRApy [5] implements the SOA approach *via* the ODE solvers provided by SciPy [6] with the function scipy.integrate.solve_ivp, which offers a wide selection of both explicit and implicit methods with various degrees of accuracy.

While biotechnology research has traditionally focused on individual microbial strains that have the necessary metabolic functions to achieve a particular task, in nature cellular species generally occur in communities in which the behavior and fitness of the individual species are interdependent. Both steady-state and dynamic FBA methods have been developed and applied to the analysis and design of community metabolism [7]. The application of dFBA to communities can be viewed as the task of assigning each species to a separate compartment, assuming that metabolites are exchanged through a shared extracellular compartment. The dFBA approach allows secreted metabolites to accumulate and affect the uptake rates of the supplied substrates and other exchanged metabolites. Complex regulatory mechanisms can be included in the exchange expressions.

There are two basic implementations of SOA dFBA for cellular communities. The first approach, called *joint* dFBA (jdFBA) [8], requires the establishment of a *community model*. The community model represents the combined stoichiometric matrices of *n* GEMs, and a *community biomass* is defined as the total sum of individual biomasses. The formation of this species is then set as the objective function.

The key idea of the second approach, called *dynamic parallel* FBA (dpFBA) [9], is to perform dFBA on individual models while keeping track of the shared *pool* of external metabolites from which all models can take up nutrients. The concentrations of each biomass and of the external metabolites is updated at each time interval.

Due to growing interest in cellular communities, and the remarkable proliferation of methods in the field, applicable tools that parallel the corresponding methods for single species have become scattered among several software packages, forcing users to a steep learning curve that includes switching often computational environments to pursue different goals. COBRApy currently offers only an out-of-the-box dFBA implementation for the simulation of the batch growth of a single species over an integration interval from $t_0$ to $t_f$, divided in *n* time steps. Here we describe a simple method to extend COBRApy dFBA to a community of species without the need for any modification of the current package.

*Method.*

Our method involves 6 steps:

1. ***k*** species in the community are represented as separate compartments with the same identical initial extracellular concentrations of metabolites at time $t_0$.

2. Given a time interval from $t_0$ to $t_f$, divided in *n* time steps, dFBA is carried out separately for **each of the *k*** compartments for only **1 time step**.

3. At the end of each time step biomass and external metabolites are updated separately in **each of the *k*** compartments.

4. Then, the concentration of external metabolites that are shared by different species are updated in **each of the *k*** compartment by adding to their values at the beginning of the iteration the changes produced in **all the *k*** compartments at the end of the iteration.

5. The next iteration of **1 time step** is carried out with updated biomasses and external metabolites.

6. Iterations continue until $t_f$ is reached or the FBA solution for **all the *k*** species becomes infeasible.

In pseudocode:

```
# Step1: represent species as separate compartments
INPUT:
   k = number of species/compartments
   n = number of time steps
   t0 = initial time
   tf = final time
   Δt = (tf - t0) / n   # time step size
   Biomass[i] = initial biomass of species i for i = 1..k
   Exchange_rates[i] = initial exchange rates for compartment i for i = 1..k
   Metabolites = initial extracellular metabolite concentrations (shared)

PROCEDURE:
   Set time = t0

   WHILE time < tf:

      # Step 2: Carry out dFBA for each species independently
      FOR i = 1 to k:
         Solve FBA for compartment i
         Compute biomass growth and metabolites changes in compartment i over Δt
         Store results:
```

```
        ΔBiomass[i], ΔMetabolites[i]

    # Step 3: Update biomass in each compartment
    FOR i = 1 to k:
      Biomass[i] = Biomass[i] + ΔBiomass[i]
      Update Exchange_rates[i] based on updated Biomass[i]

    # Step 4: Update shared extracellular metabolites
    FOR each metabolite m:
      Metabolites[m] = Metabolites[m] + Σ(ΔMetabolites[i][m]) over all i

    # Step 5: Advance to next time step
    time = time + Δt

    # Step 6: Check feasibility
    IF FBA solution infeasible for ALL species:
      EXIT loop

  END WHILE

OUTPUT:
  Time-course trajectories of Biomass[i] and Metabolites
```

Since at each iteration the concentration of external metabolites in the shared pool is updated independently for each one of the $k$ compartments, the changes in shared metabolites produced by the other compartments during the $\Delta t$ covered by the iteration are ignored. Although, before the next iteration the concentration of shared metabolites is updated taking into consideration the contribution of all the compartments, the individual updates introduce an integration error that make the procedure again akin to a *Forward Euler* method with 1st order accuracy $O(\Delta T)$. Therefore, a very small $\Delta T$ must be taken to ensure the result is as correct as possible.

To evaluate the magnitude of the discretization error introduced by this method we have first used as reference the simple tutorial on dFBA (https://cobrapy.readthedocs.io/en/latest/dfba.html) included in the COBRApy documentation. In this tutorial, using a Core Metabolism model of *E. coli*, glucose consumption and biomass growth were simulated over an interval of 8 hours using the BDF option of the ODE solver. The BDF option calls an *implicit multi-step variable-order stiff method*. A problem is *stiff* if some parts of the solution vary slowly, but other parts vary rapidly, so that the numerical method must take small steps to obtain satisfactory results. Although optimized, a constant stepsize may not be appropriate for the entire range of integration: for example, chemical reactions typically start off with rapid changes before becoming smooth. This behaviour is usually described as *stiffness*. The panels of Figure 1 shows superimposed the changes in glucose and biomass concentrations calculated using the original COBRApy protocol with BDF solver, and those calculated using our discretized method with the ODE solver option RK45 and different integration stepsizes ($\Delta t = 1.0, 0.2, 0.05, 0.01\ hrs$). RK45 is an adaptive *explicit Runge-Kutta non-stiff method of order 5(4)*, in which the error is controlled assuming accuracy of the fourth-order method, but steps are taken using the fifth-order accurate formula. The *stiff* behavior of the *BDF* solver can be recognized in the uneven spacing of the integration points calculated using the original protocol. Since our method is currently based on a constant small stepsize to update the external metabolites, we have found to be computationally more efficient to use the *non-stiff* RK45 solver which is much faster (per step size) than the *stiff* BDF solver (although the latter may use fewer points).

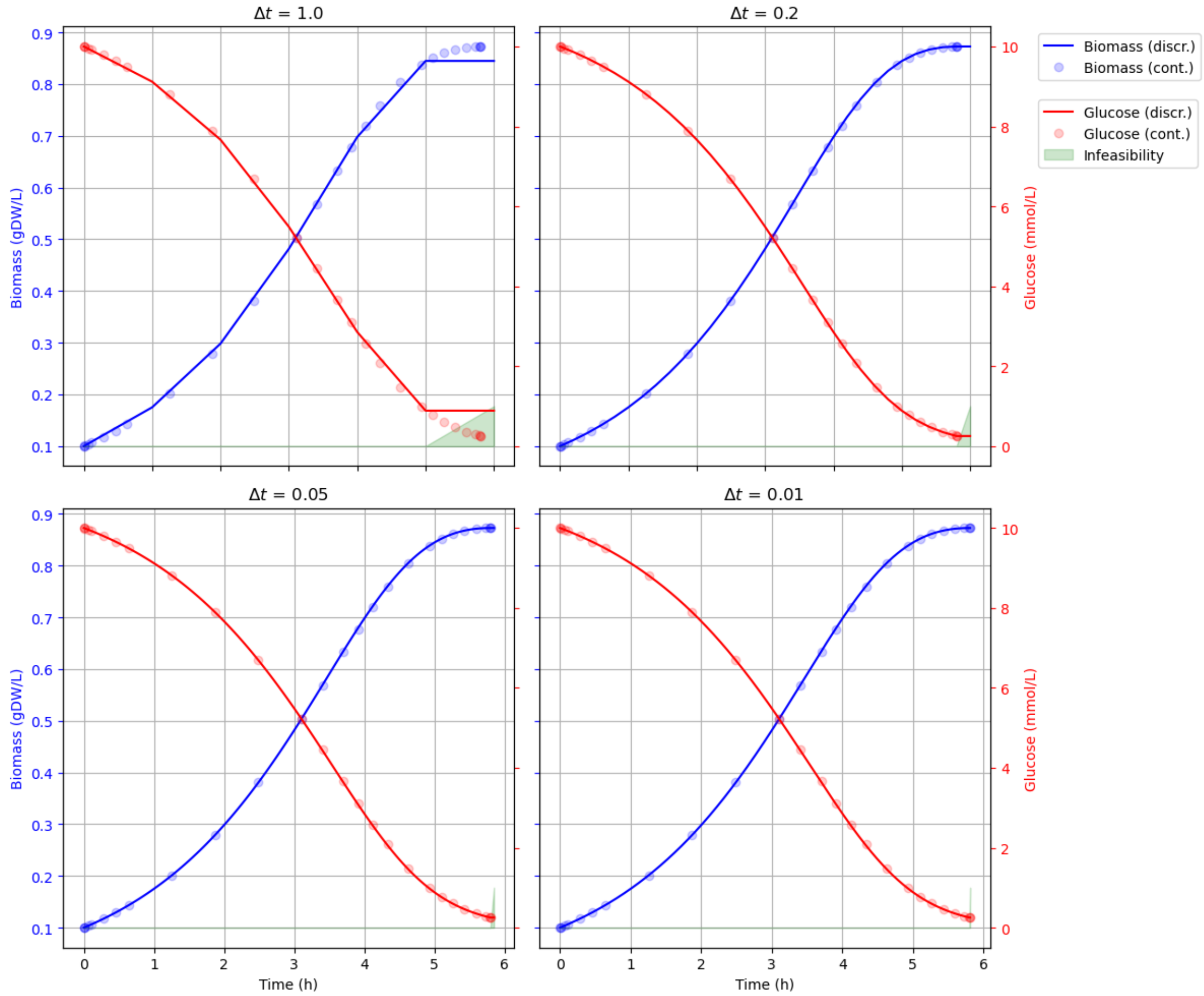

**Figure 1**. Panels show superimposed the changes in glucose and biomass concentrations calculated using the original COBRApy dFBA protocol with BDF solver (shaded circles), and those calculated using the discretized dFBA method (continuous lines) with RK45 solver and different integration stepsize, $\Delta t = 1.00, 0.20, 0.05, 0.01$ h. A shaded green area displays the part of the time span in which the solution provided by the discretized method becomes infeasible.

This initial test shows clearly that the discretization error can be progressively reduced by choosing a sufficiently small stepsize. In fact, in this case a still rather large stepsize of 0.2 hrs already provides an acceptably low error in the overall integration.

One convenient feature of the discretized dFBA method described here is that an ensemble of species can be easily simulated in parallel, each species model utilizing its own objective function and optimization protocol (for example, we could use *parsimonious* or even *geometric* rather than *standard* FBA), and its own pool of external metabolites during the integration $\Delta t$ step size. At the end of each iteration, only the concentration of the metabolites shared by two or more species is updated in the pool.

To evaluate this feature of the method, in a second test we sought to simulate the growth of two mutant strains of *E. coli*, one with a respiratory defect and one with a fermentation defect, in a bioreactor, at an initial [glucose] = 20 mM, and constant [O2] = 1.0 mM. In this case, the respiratory deficient strain utilizes glucose from the shared pool of external metabolites to actively secrete acetate and ethanol. At the same time the fermentation deficient strain uses acetate and ethanol secreted by the first strain in the shared pool as respiratory substrates. While oxygen and ethanol diffuse freely through the membrane, we still represented $O_2$ consumption by the *E. coli* terminal oxidase as a saturable process. The stepsize was 0.01 hrs. The outcome of this test is shown in Figure 2.

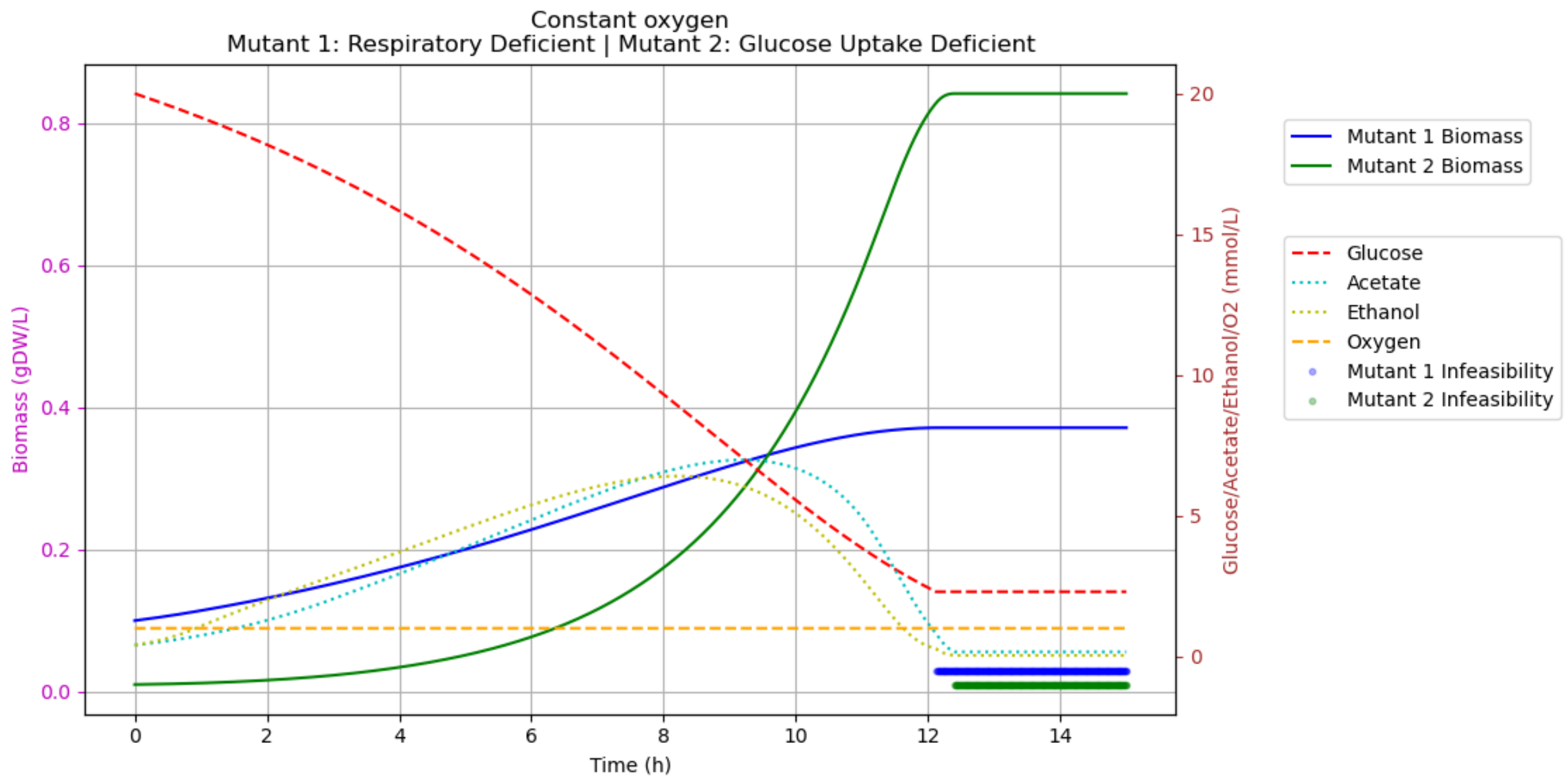

**Figure 2**. Metabolites and biomass concentrations changes for the collaborative growth of a respiration deficient strain and a fermentation deficient strain of *E. coli* in bioreactor with constant $O_2$ concentration. Blue and green bars display the part of the interval in which the FBA solution becomes infeasible for each model.

In this case we see that while the fermentation deficient strain (Mutant 2) starts from a much smaller biomass, its utilization of acetate and ethanol produced by the respiratory deficient strain (Mutant 1) allows it to ultimately take over the culture by the time all the glucose is used.

In a second test (Figure 3), we simulated the condition in which both mutants grow in a bioreactor with constant [Glucose] = 20 mM and [$O_2$] = 1 mM concentrations. Since in this simulation, the substrates are constantly replenished, we followed the growth of the two mutants for a much longer time (30 hrs). The stepsize was 0.005 hrs. The result of this simulation is shown in Figure 3.

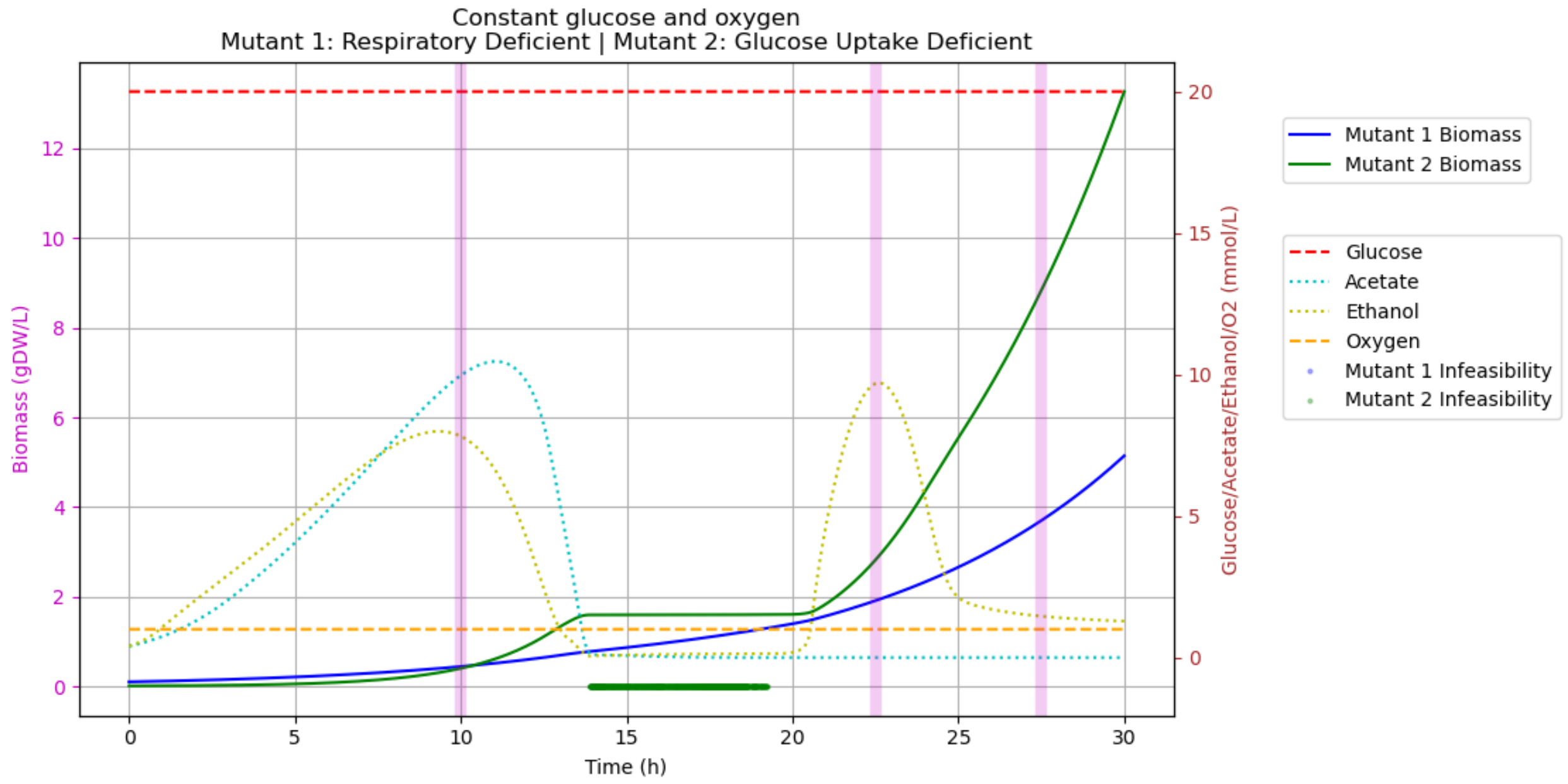

**Figure 3**. Metabolites and biomass concentrations changes for the collaborative growth of a respiration deficient strain and a fermentation deficient strain of *E. coli* in a bioreactor with constant [glucose] = 20.0 mM and [$O_2$] = 1 mM concentration. Blue and green bars display the part of the interval in which the FBA solution becomes infeasible for each model. Shaded vertical bars indicate three time points (10, 22.5, 27.5 hrs) at which *species specific* fluxes for the shared metabolites were examined (Table 1).

| **Table 1** | hour | Glucose | Acetate | Ethanol | O2 | Biomass (μ) |
|---|---|---|---|---|---|---|
| Mutant 1 | 10.0<br>22.5<br>27.5 | -7.999<br>-7.994<br>-7.988 | 6.933<br>0.000<br>0.000 | 6.774<br>13.703<br>13.595 | 0.0<br>0.0<br>0.0 | 0.151<br>0.132<br>0.132 |
| Mutant 2 | 10.0<br>22.5<br>27.5 | 0.0<br>0.0<br>0.0 | -5.578<br>0.0<br>0.0 | -8.864<br>-9.052<br>-5.471 | -19.794<br>-14.242<br>-9.278 | 0.411<br>0.295<br>0.163 |

**Table 1.** External metabolites fluxes (negative for uptakes and positive for export) are in unit of *mmol/gDW/h*. Biomass growth rate (μ) is in unit of $h^{-1}$.

Past the 13$^{th}$ hour the fermentation deficient Mutant 2 displays short periods of infeasibility without growth, during which it waits for the respiratory deficient Mutant 1 to export enough ethanol/acetate to support again its metabolic needs.

Since we save all the species fluxes at every point in the simulation, we can visualize the metabolic changes that occur in each species. The four panels below display Escher maps of Mutant 1 and Mutant 2 at 10 and 27 hrs, respectively. For example, we can see that Mutant 1 wastes some glucose to produce formate that, at least in our dynamic set up, cannot be used by Mutant 2.

MUTANT 1: 10 hrs

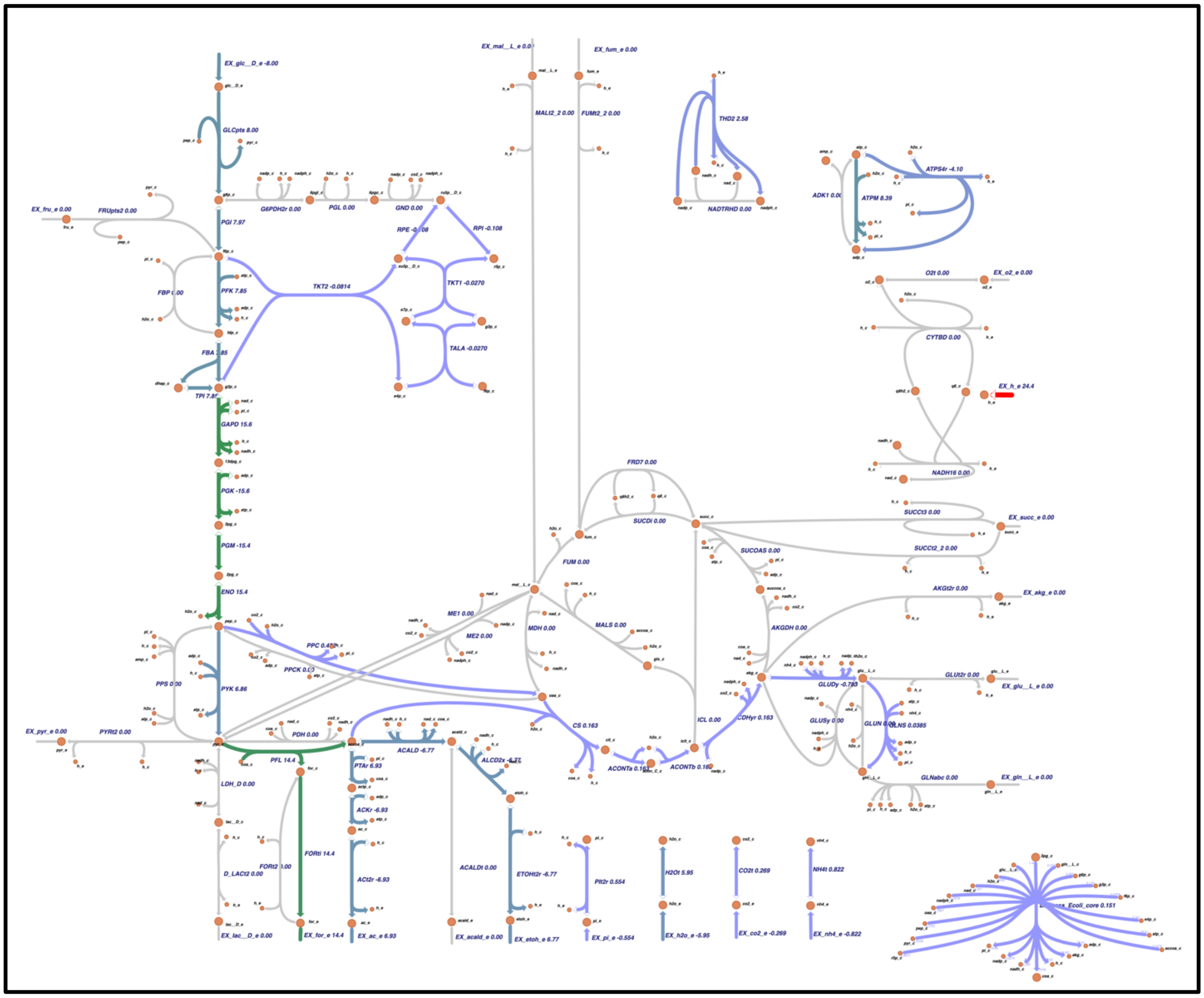

MUTANT 2: 10 hrs

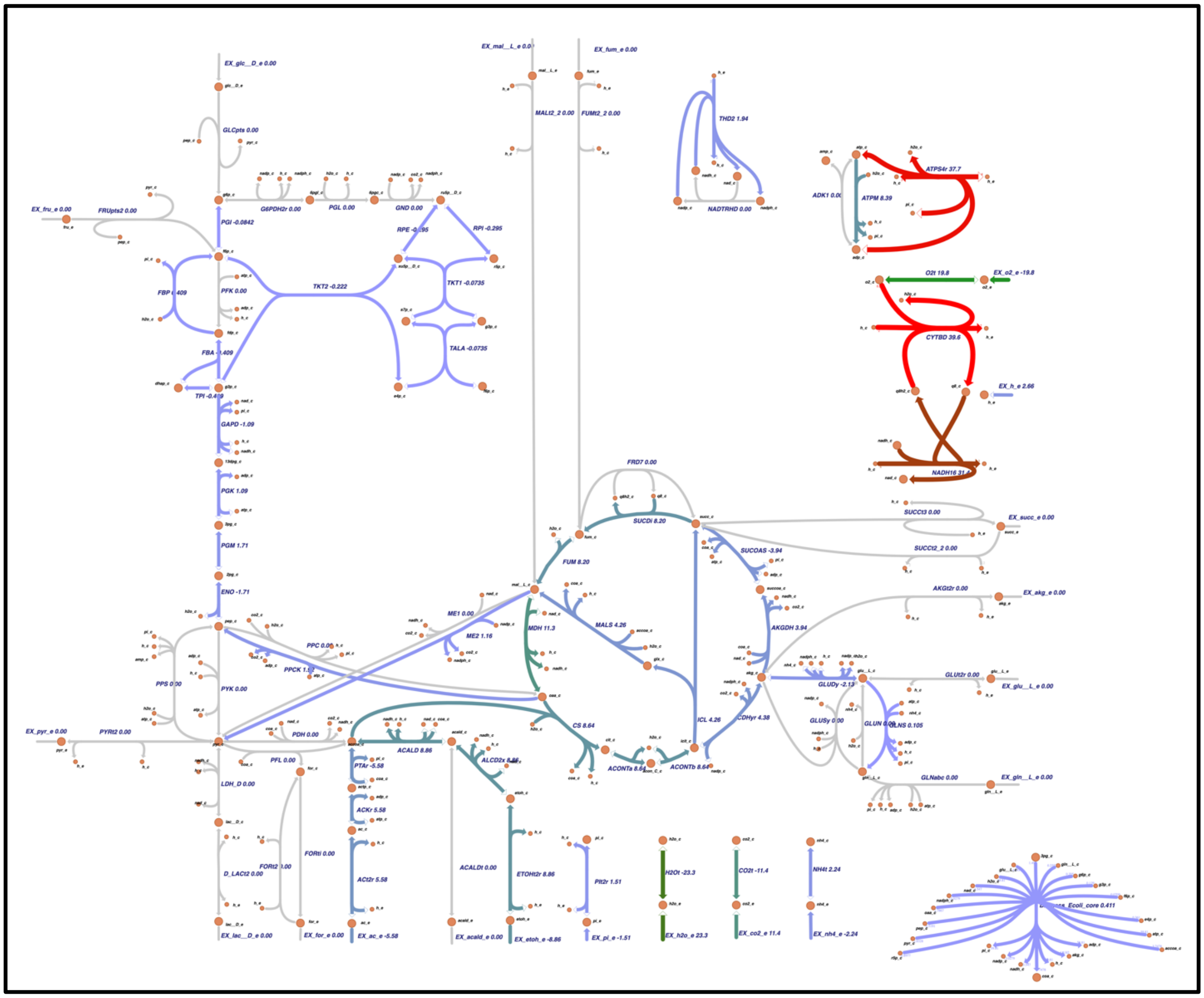

MUTANT 1: 27.5 hrs

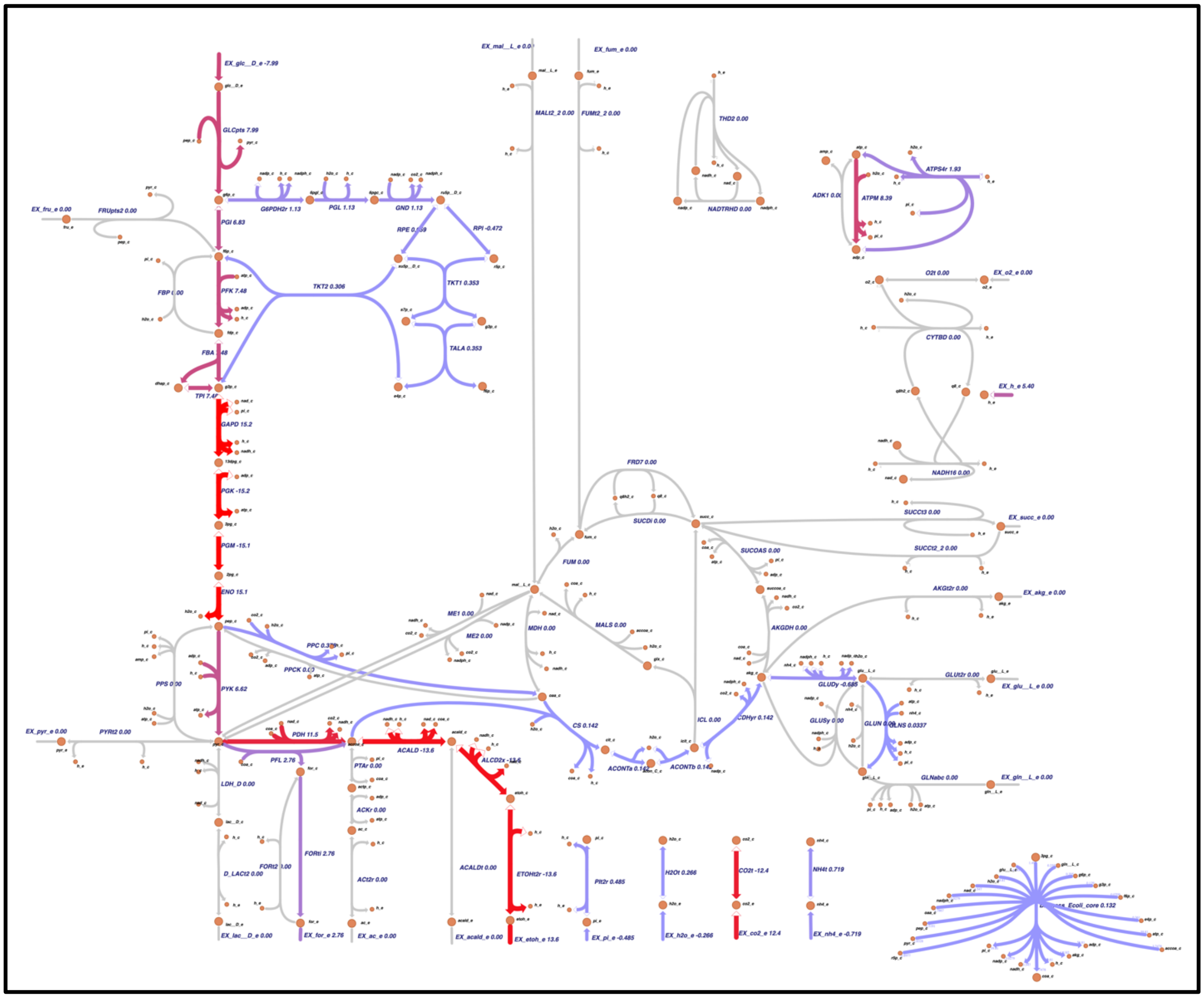

MUTANT 2: 27.5 hrs

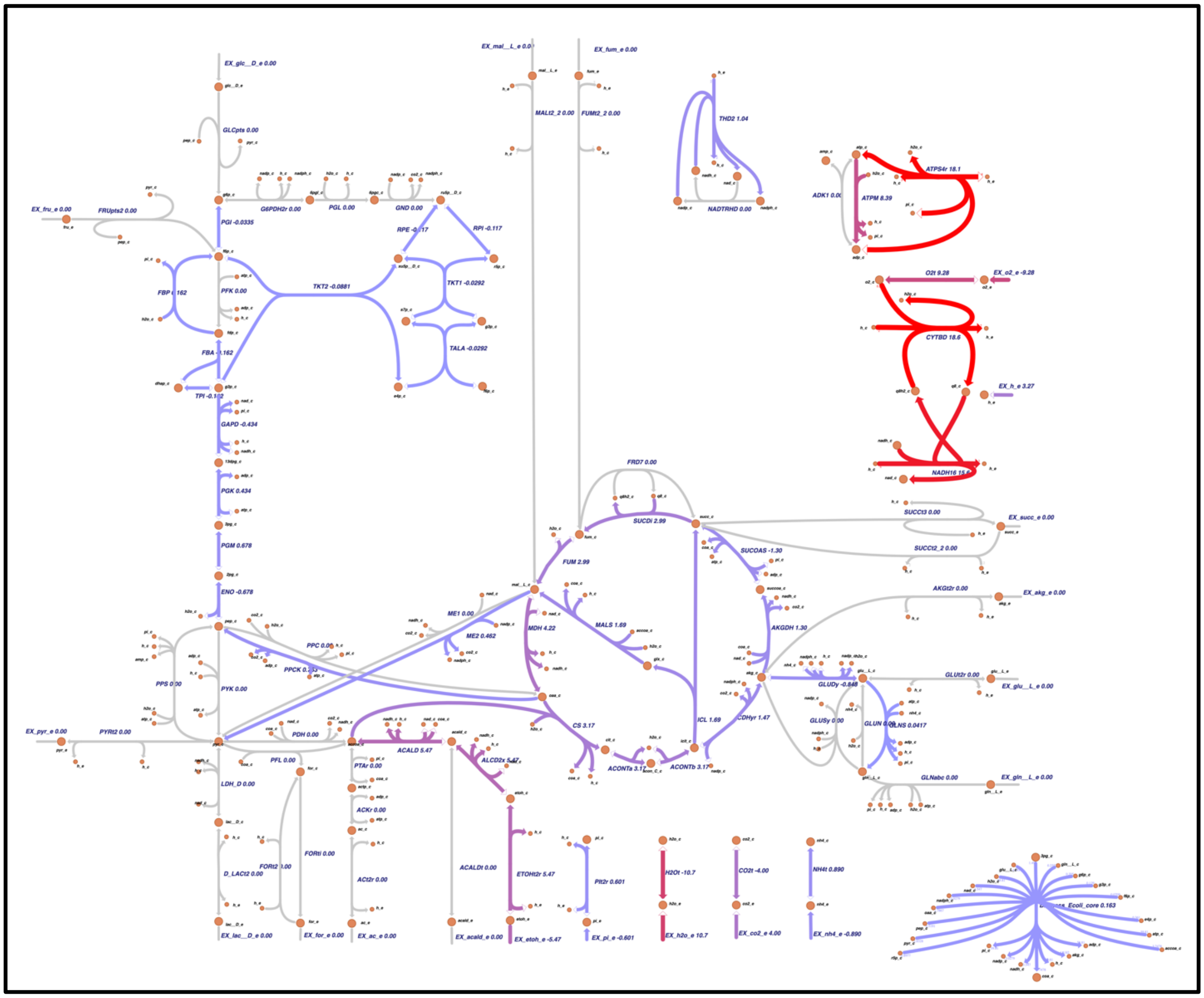

It should be noticed that while in the tests of Figures 2 and 3 the concentration of glucose or/and $O_2$ were kept constant, the fact that the integration of the entire interval takes place in a customizable number of steps, makes it possible to develop a very flexible schedule to change the environment at every step or group of steps. For example, rather than having a constant concentration of a substrate, we could update it every *n* iterations, or only prevent its concentration from falling below a given minimum.

*Conclusion.*

In this report we show that the existing dFBA implementation included in COBRApy for the growth of a single species can be easily extended to the simulation of the time evolution of a community of species using a recursive discretization strategy in which at the end of each iteration the environment metabolites and biomass changes are updated independently for each species, and at the beginning of the next iteration the concentration of the shared metabolites is updated to incorporate the changes produced by all the individual updates. This strategy allows for the added flexibility of including in the community models of different levels of complexity and different objectives for the individual species, as well as the opportunity to change objective at each iteration. The initial individual updating of the concentrations of external metabolites at the end of each iteration can be intuitively interpreted as reflecting the changes of each species microenvironment, before these changes propagate to a wider shared environment. Expressions representing more explicitly this phenomenon can be introduced at each iteration. By the same token, simulation of communities of many different cell types may require adopting very small $\Delta t$ stepsizes to avoid that the accumulated changes produced by a large number of compartments at the beginning of each iteration produce excessive fluctuations of the external metabolites.

*Code availability.*

COBRApy code to reproduce all test cases is available at:
https://github.com/dgattiwsu/dpFBA-with-COBRApy/tree/main/tests.

*References.*

[1] B. O. Palsson, *Systems Biology: Constraint-Based Reconstruction and Analysis.* Cambridge University Press., 2015.

[2] A. Bordbar, J. M. Monk, Z. A. King, and B. O. Palsson, "Constraint-based models predict metabolic and associated cellular functions," *Nat Rev Genet,* vol. 15, no. 2, pp. 107-20, Feb 2014, doi: 10.1038/nrg3643.

[3] R. Mahadevan, J. S. Edwards, and F. J. Doyle, 3rd, "Dynamic flux balance analysis of diauxic growth in Escherichia coli," *Biophys J,* vol. 83, no. 3, pp. 1331-40, Sep 2002, doi: 10.1016/S0006-3495(02)73903-9.

[4] J. C. Butcher, *Numerical Methods for Ordinary Differential Equations*. Wiley, 2016, p. 544.

[5] A. Ebrahim, J. A. Lerman, B. O. Palsson, and D. R. Hyduke, "COBRApy: COnstraints-Based Reconstruction and Analysis for Python," *BMC Syst Biol,* vol. 7, p. 74, Aug 8 2013, doi: 10.1186/1752-0509-7-74.

[6] P. Virtanen *et al.*, "SciPy 1.0: fundamental algorithms for scientific computing in Python," *Nat Methods,* vol. 17, no. 3, pp. 261-272, Mar 2020, doi: 10.1038/s41592-019-0686-2.

[7] A. Heinken, A. Basile, J. Hertel, C. Thinnes, and I. Thiele, "Genome-Scale Metabolic Modeling of the Human Microbiome in the Era of Personalized Medicine," *Annu Rev Microbiol,* vol. 75, pp. 199-222, Oct 8 2021, doi: 10.1146/annurev-micro-060221-012134.

[8] E. Tzamali, P. Poirazi, I. G. Tollis, and M. Reczko, "A computational exploration of bacterial metabolic diversity identifying metabolic interactions and growth-efficient strain communities," *BMC Syst Biol,* vol. 5, p. 167, Oct 18 2011, doi: 10.1186/1752-0509-5-167.

[9] S. Stolyar *et al.*, "Metabolic modeling of a mutualistic microbial community," *Mol Syst Biol,* vol. 3, p. 92, 2007, doi: 10.1038/msb4100131.